\documentclass{webofc}
\usepackage[varg]{txfonts}
\usepackage[utf8]{inputenc}
\usepackage{lineno}
\usepackage{svg}
\usepackage{afterpage}
\usepackage{hyperref}
\usepackage{caption}
\usepackage{siunitx}
\usepackage{subcaption}
\usepackage{natbib}

\newcolumntype{Y}{>{\centering\arraybackslash}X}

\newcommand{\orcid}[1]{\href{https://orcid.org/#1}{\includesvg[width=10pt]{figs/orcid.logo.icon.svg}}}

\graphicspath{{figs/}}

\title{Machine learning for surface prediction in ACTS}
\date{\today}

\begin{document}

\author{
    \firstname{Benjamin} \lastname{Huth} \inst{1} \orcid{0000-0002-3163-1062}\and
    \firstname{Andreas} \lastname{Salzburger} \inst{2} \orcid{0000-0001-6004-3510}\and
    \firstname{Tilo} \lastname{Wettig} \inst{1}
}

\institute{
Department of Physics, University of Regensburg, 93040 Regensburg, Germany\and
CERN, Esplanade des Particules 1, 1211 Meyrin, Switzerland
}

\abstract{We present an ongoing R\&D activity for machine-learning-assisted navigation through detectors to be used for track reconstruction. We investigate different approaches of training neural networks for surface prediction and compare their results. This work is carried out in the context of the ACTS tracking toolkit.}

\maketitle

%%%%%%%%%%%%%%%%%%%
\section{Introduction}\label{sec:intro}
	
	Track reconstruction is a challenging part of event reconstruction in HEP experiments. It is usually CPU intensive and often prone to edge cases that need carefully crafted solutions  for efficient and reliable results. One main task during track reconstruction is the transport of the track parameterization (i.e., the representation of the trajectory, either in free or in surface-bound coordinates) to another point in the detector. While this transport may have analytic solutions for certain special cases, in the general case it needs the numerical solution of the equations of motion through the detector setup. Such a general solution is implemented as a fourth-order Runge-Kutta numerical integrator in the ACTS tracking toolkit~\cite{acts}. In a real-life experiment, however, the material inside the detector causes additional changes and uncertainties in the track propagation, manifested in a change of momentum due to energy-loss effects and to additional noise terms in the covariance matrix caused by multiple Coulomb scattering and energy-loss straggling. In order to adequately apply such correction terms, the integrator must be aware of its position in the detector geometry. For this task a navigation module is required.
	
	The environment in which this navigation takes place is built up from individual surfaces, which are either sensitive surfaces or volume boundary surfaces. The navigation problem consists of predicting the next surface the track will cross in the detector geometry.
	
\section{Motivation: limitations of the current ACTS navigator}\label{sec:limitations}
    
    A modern particle detector has thousands of individual measurement surfaces, e.g., silicon sensors in the pixel and/or strip detectors. In order to allow for acceptable timing performance when navigating through the detector, a selection of candidate surfaces has to be done. In ACTS this is realized through a dedicated navigator module that suggests the most likely next candidate to the numerical integrator, which then performs the solution of the differential equation of motion. The navigator module builds upon a hierarchical structure of volumes, layers and grid structures in which the candidate surfaces are filled and from which they are retrieved when the track is in local proximity.
    
    While such an approach has been proven to be very effective, it comes at a price: geometry building and presorting is highly tailored to a specific detector layout and needs a certain amount of tuning to find the sweet spot between timing and navigation performance. The code for this kind of navigation is highly branched and complex, as it tries to follow a logical hierarchy tree through the detector. Attempts to apply the same code, e.g., on accelerated hardware have failed so far for several reasons: the geometry classes for guiding through the detector are highly polymorphic, and dynamic branches make large-scale parallelization difficult. In addition, unconventional detector layouts require dedicated modules that might not exist yet within the ACTS toolkit. Finally, this solution is not free from errors.
    
\section{ML-assisted navigation}

    The navigation module is not required to give an exact prediction of the next surface to be intersected by the trajectory. Rather, it provides an inclusive list of candidate surfaces which are then attempted to be intersected. In the following, a machine-learning (ML) approach to finding these candidate surfaces is presented. This approach has the potential to overcome the limitations described in Sec.~\ref{sec:limitations}.

    \subsection{Training data}
	    The underlying idea of ML-assisted navigation is that the necessary information to navigate through the detector can be learned from tracks simulated in this specific detector setup. In principle a navigator is needed beforehand to be able to generate these data. In the case of the results presented in this paper, the stable and well-tested infrastructure of ACTS is used for this.
	    
	    Even in the absence of an existing navigation module, an ML-assisted navigation can be applied: the navigation problem is inherently deterministic, and a brute-force trial-and-error navigator will --- together with the cross-check through the numerical integration --- provide a precise solution to the problem. This is computationally very inefficient but only needs to be done once to train a navigator, which will then be much more efficient afterwards.
	
	\subsection{Representing surfaces}
		The basic entities in this context are \emph{surfaces}. The first question that arises is how a surface can be represented in the training data so that a neural network can process it properly. In ACTS each surface has a unique identifier, \texttt{Acts::GeometryId} (in fact just a wrapper around an integer). This integer contains information about the surface that can be decoded by applying certain bit masks to the integer. However, using the numerical value of this integer in computations is not sensible because the results of operations such as addition or multiplication have no interpretation. Rather, this integer represents a \emph{categorical variable}.
		
		There exist different approaches to handle categorical variables in neural networks. One way to pass categorical values to a neural network is to use a \emph{one-hot encoding}. This means that each possible category is represented by an entry in a $ d $-dimensional vector, where $ d $ is the number of categories. The entry of the correct category is set to 1, all others to 0. However, since there are several thousand surfaces, the high dimension of the feature space would cause both algorithmic problems (\emph{curse of dimensionality}) and computational inefficiencies (because the data are extremely sparse). Furthermore, one-hot-encoded values only allow one to distinguish categories but are not able to represent more complex relations between them.
		
		An alternative approach that is capable of representing such relations is the use of a technique called \emph{embedding}. Each category is mapped to a point in an $n$-dimensional space. In this space, the relative positions of these points can encode relations between the categories. This technique was successfully applied to different fields in recent years \cite{embeddings_rossmann_challenge}. In the case of surfaces, either an $n$-dimensional embedding can be trained or simply the 3D center coordinates of the surfaces can be used. Other options to generate embeddings are combinations of the above-mentioned approaches or the application of techniques such as \emph{metric learning}. However, these were not considered in the work presented here.
		
	    Representing each surface as a point in an embedding space works fine for most surfaces in the detector. However, if a surface has a large spatial extent, tracks crossing it can have a wide variety of target surfaces, and it may be more difficult to predict the next surface accurately. In the tested setup, this applies mainly to the \emph{beam-pipe surface}, an innermost cylinder that represents the actual beam containment and extends along the entire detector. In the training data, this one surface represents all track origins (while in reality, the tracks originate on the beam line). A possible optimization here is to split this surface into different sections along the beam-pipe axis, and generate an embedding for each of them. One can then even further split the track origins depending on the polar angle of their unit direction, $ \phi = \mbox{arctan2}(d_x, d_y) $. The splitting in the $ z $-direction is performed such that the number of track origins per bin is roughly the same (as a consequence, their boundaries are not equidistant on the $ z $-axis). The beam-pipe split will be denoted as a tuple of two integers in the following: $ (z$-$\text{split}, \phi$-$\text{split}) $. For certain models the use of beam-pipe splitting can improve the prediction performance at the beam pipe significantly.
		
	\subsection{Predicting the next surface}\label{sec:approaches}
		The problem setup for predicting the next surface is the following. The starting point is always on a surface, and at this point all information about the current track (position, momentum, charge) is available.\footnote{Note that the prediction does not attempt to estimate the full track state at the next surface. This must be done by the integration component of the particular propagator implementation.} Given this information, the next surface the track will cross must be predicted. In the following, two different approaches to this problem are considered:
		
		\begin{figure}
			\centering
			\begin{subfigure}[t]{0.49\linewidth}
				\centering
				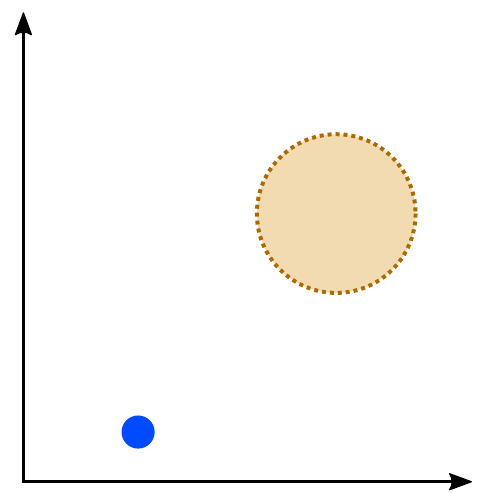
				\caption{Target-prediction approach: Given the embedding $ \vec{s}_i $ of the starting surface and the track parameters $ \vec{p}_i $, a target embedding $ \vec{s}_f $ is predicted which is then resolved to an actual surface by nearest-neighbor search.}
				\label{fig:target_pred}
			\end{subfigure}
			\hfill
			\begin{subfigure}[t]{0.49\linewidth}
				\centering
				%% Creator: Inkscape 1.0.2 (e86c870879, 2021-01-15), www.inkscape.org
%% Accompanies image file 'pairwise_score_scheme.pdf' (pdf, eps, ps)
\begingroup%
  \makeatletter%
  \providecommand\color[2][]{%
    \errmessage{(Inkscape) Color is used for the text in Inkscape, but the package 'color.sty' is not loaded}%
    \renewcommand\color[2][]{}%
  }%
  \providecommand\transparent[1]{%
    \errmessage{(Inkscape) Transparency is used (non-zero) for the text in Inkscape, but the package 'transparent.sty' is not loaded}%
    \renewcommand\transparent[1]{}%
  }%
  \providecommand\rotatebox[2]{#2}%
  \newcommand*\fsize{\dimexpr\f@size pt\relax}%
  \newcommand*\lineheight[1]{\fontsize{\fsize}{#1\fsize}\selectfont}%
  \ifx\svgwidth\undefined%
    \setlength{\unitlength}{141.73228346bp}%
    \ifx\svgscale\undefined%
      \relax%
    \else%
      \setlength{\unitlength}{\unitlength * \real{\svgscale}}%
    \fi%
  \else%
    \setlength{\unitlength}{\svgwidth}%
  \fi%
  \global\let\svgwidth\undefined%
  \global\let\svgscale\undefined%
  \makeatother%
  \begin{picture}(1,1)%
    \lineheight{1}%
    \setlength\tabcolsep{0pt}%
    \put(0,0){\includegraphics[width=\unitlength,page=1]{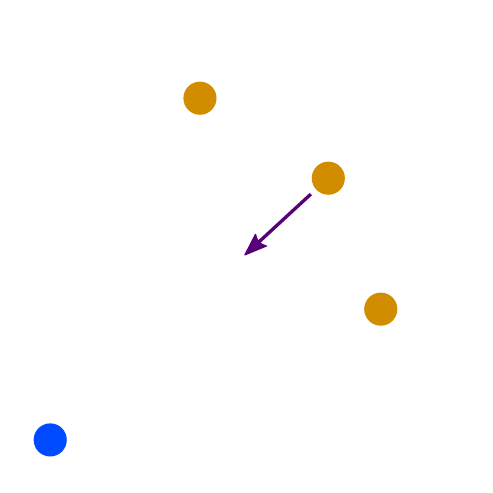}}%
    \put(0.38636208,0.86216346){\color[rgb]{0,0,0}\makebox(0,0)[lt]{\lineheight{1.25}\smash{\begin{tabular}[t]{l}$ \large \color[HTML]{D18C00} \boldsymbol{ \vec{s}_a } $\end{tabular}}}}%
    \put(0.64149597,0.69774384){\color[rgb]{0,0,0}\makebox(0,0)[lt]{\lineheight{1.25}\smash{\begin{tabular}[t]{l}$ \large \color[HTML]{D18C00} \boldsymbol{ \vec{s}_b } $\end{tabular}}}}%
    \put(0.75299896,0.42484501){\color[rgb]{0,0,0}\makebox(0,0)[lt]{\lineheight{1.25}\smash{\begin{tabular}[t]{l}$ \large \color[HTML]{D18C00} \boldsymbol{ \vec{s}_c } $\end{tabular}}}}%
    \put(0.17154772,0.06690155){\color[rgb]{0,0,0}\makebox(0,0)[lt]{\lineheight{1.25}\smash{\begin{tabular}[t]{l}$ \large \color[HTML]{004BFF} \boldsymbol{ \vec{s}_i, \vec{p}_i } $\end{tabular}}}}%
    \put(0,0){\includegraphics[width=\unitlength,page=2]{pairwise_score_scheme.pdf}}%
  \end{picture}%
\endgroup%

				\caption{Score-prediction approach: The embedding $ \vec{s}_i $ of the starting surface, the embedding of a possible target surface (e.g., $ \vec{s}_a $) and the track parameters $ \vec{p}_i $ are used to predict the probability that this candidate target is the actual target. }
				\label{fig:pairwise_score}
			\end{subfigure}
			\caption{Visualization of the two surface-prediction methods.}
		\end{figure}
		
		\begin{enumerate}
			\item \textbf{Predict target directly.} The current surface $ \vec{s}_i $ (represented as a vector in the embedding space) and the track parameters $ \vec{p}_i $ are fed into the neural network that then predicts a point $ \vec{s}_f $ in the embedding space:
			\begin{linenomath*}\begin{equation}
					(\vec{s}_i, \vec{p}_i) \rightarrow \vec{s}_f\,.
			\end{equation}\end{linenomath*}
			Since this point will hardly ever match exactly any representation of another surface, it is resolved to $ k $ candidate surfaces with a \emph{k-nearest-neighbor search} (KNN search) in the embedding space (see Fig.~\ref{fig:target_pred}). 
			
			This approach has the advantage that it can predict the target without any candidates known beforehand. However, a nearest-neighbor search is a computationally very expensive operation.
			
			\item \textbf{Predict score of surface pair:} In this case, a fixed set of candidate target surfaces is given for each starting surface. Together with the starting surface they effectively form a graph of all considered paths through the detector. The representation of the starting surface $ \vec{s}_i $, the representation of a potential target surface $ \vec{s}_f $ and the track parameters $ \vec{p}_i $ are fed into the network. The result is a number between 0 and 1 representing how likely the network considers the tested candidate surface to be the next surface:
			\begin{linenomath*}\begin{equation}
					(\vec{s}_i, \vec{s}_f, \vec{p}_i) \rightarrow \text{score}\in(0,1)\,.
			\end{equation}\end{linenomath*}
			Then all candidates are sorted by score and used for navigation (see Fig.~\ref{fig:pairwise_score}). 
			
			The advantage of this method is that it works with a fixed set of surface candidates. Since no nearest-neighbor search needs to be performed, the method is computationally less expensive. On the other hand, the method can only find surfaces that are present in the graph. Hence it can happen that the correct surface cannot be found, in which case a fallback solution must be implemented. If this happens rarely enough, the score-prediction method is a good option.
		\end{enumerate}

\section{Training process and test results}
	
	In this section, the setup for the different approaches is described and their results are presented and discussed.

	\subsection{Training-data generation, model architecture and software environment}
		The data used for training, validating and testing the navigation models are generated using the ACTS propagation example (settings: a flat distribution in momentum between 0.1 and \SI{100}{\giga\electronvolt} and angular uniform distributions within detector acceptance) with a custom logger to record the necessary data. All results shown in this paper are produced with the data of roughly 500 thousand simulated tracks in ACTS' \emph{generic detector} that has also been used for the Tracking Machine Learning Challenge (TrackML) \cite{track_ml_challenge}. Two thirds of these data are used for training, and the remaining third is used for evaluation after the training. The data are split on a track basis, so only complete tracks are present in the training and evaluation datasets. The pre-trained embeddings (see next subsection) were fitted using a set of 128 thousand tracks. 
		
		All presented models are simple fully-connected feed-forward neural networks. As input features, surface embedding and track parameters are used as described in Sec.~\ref{sec:approaches}. The track parameters handed over to the models are the three components $ d_x, d_y, d_z $ of the unit direction and the ratio $ q/p $ of charge and magnitude of momentum. The hyperparameters used are listed in Table~\ref{tab:hyperparams}.
		
		The training and evaluation environment is based on Tensorflow/Keras 2.3.0~\cite{keras, tensorflow} on top of Python 3.8.6. All code used to generate these results can be found at~\cite{my_github_repo}.
		
		\begin{table}
			\centering\footnotesize
			\begin{tabular}{c|c|c|c|c|c}
				& \# of layers & Units per layer & Embedding & BP split $ (z, \phi) $ & loss \\
				\hline
				Target pred. & 3 & 500 & 10D, pre-trained & (40,1) & MSE \\
				\hline
				Score pred. & 4 & 750 & 3D, real space & (400,16) & binary crossentropy \\
			\end{tabular}
			\caption{Hyperparameters of the models presented. All models where trained with the help of the \emph{Adam} optimizer and a learning rate of 0.001. \emph{BP split} is short for beam-pipe split in this context.}
			\label{tab:hyperparams}
		\end{table}
	
	\subsection{Pre-training of embeddings}
		When training an embedding, there are two choices: one can train the embedding as part of the final navigation model, or one can train it beforehand and use the embedding later as input to the model.
		The first approach has the obvious advantage that the embedding can be fitted better to the specific problem but has the disadvantage of combining two different training objectives in one model. Whereas the navigation model should generalize as well as possible beyond the training data, the embedding should be fitted as well as possible to the training data, with no need for generalization. Because of these considerations and insights from numerical experiments, pre-trained embeddings were used for all results presented.
		
		Within the Keras framework, embeddings of the form $ n \in \left[0, N\right) \rightarrow \vec{v} \in \mathbb{R}^d $ (where $ n $ is an integer identifier for a specific surface and $ N $ is the number of surfaces) can be easily created and trained using the provided embedding layer. As training data, pairs of surface identifiers $ n_1 $ and $ n_2 $ with the corresponding truth information were used: $ 0 $ if they are not connected by any track in the simulation data, $ 1 $ if they are. To map the two resulting embedding vectors to a single scalar, a dot product $ \vec{v}_1 \cdot \vec{v}_2 $ is used. The dot product is then shifted and scaled by a dense layer of size 1 to match the output range $ (0,1) $. Afterwards, the embedding layer can be used to preprocess the data before feeding them into the navigation model. The embeddings used for this work are trained for 20 thousand epochs with simulated data consisting of 128 thousand tracks.
	
	\subsection{Baseline comparison: weighted-graph navigator}
		To allow for better interpretation of performance results, the performance of a very basic navigation model is used for comparison. This model always predicts the paths based on their number of occurrences in the training data (``weighted-graph navigator''). For example, the path that occurs most is always the first prediction. Even though it is only used as a performance baseline here, this approach could be used for a ``first guess'' in a real implementation, in order to save computing time in some cases. However, in a more realistic detector setup the performance of this approach is expected to deteriorate since there will be more paths available.
				
    	\begin{figure}
    		\includegraphics[width=\textwidth]{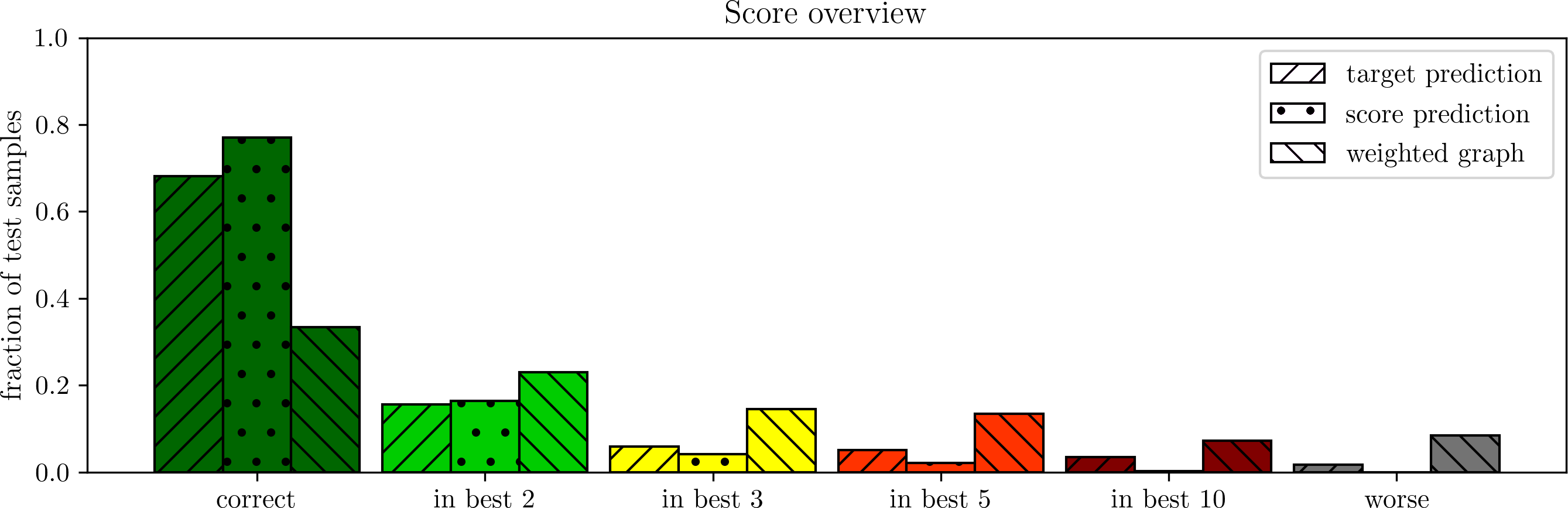}
    		\caption{Score distributions in the generic detector for the two navigation models and a weighted-graph approach for a baseline comparison.}
    		\label{fig:score_distributions}
    	    \centering
			\vspace{1mm}  % there is really just 1mm space left before warning appears
			\begin{subfigure}[t]{\linewidth}
				\centering
				\includegraphics[height=50mm]{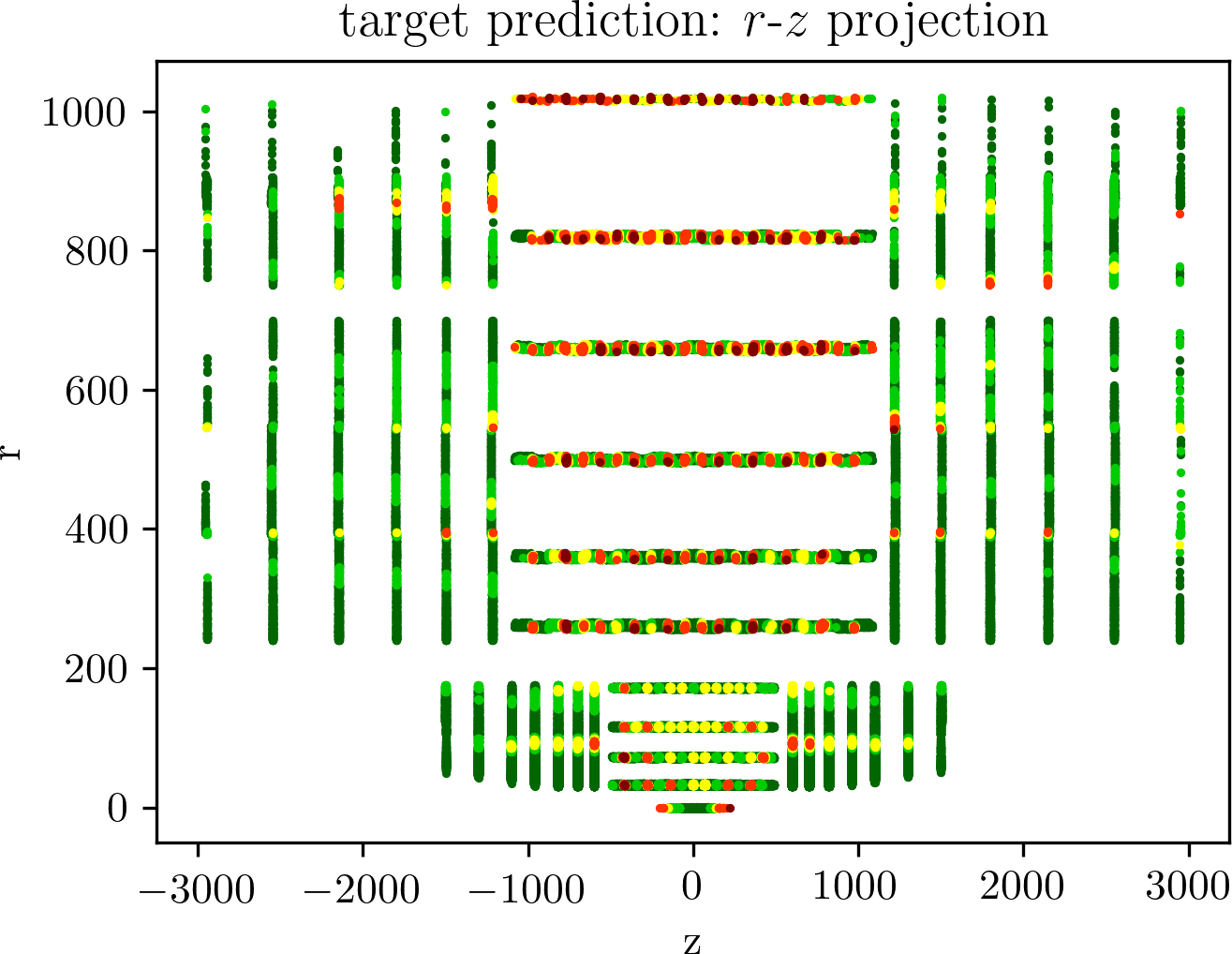}
				\hfill
				\includegraphics[height=50mm]{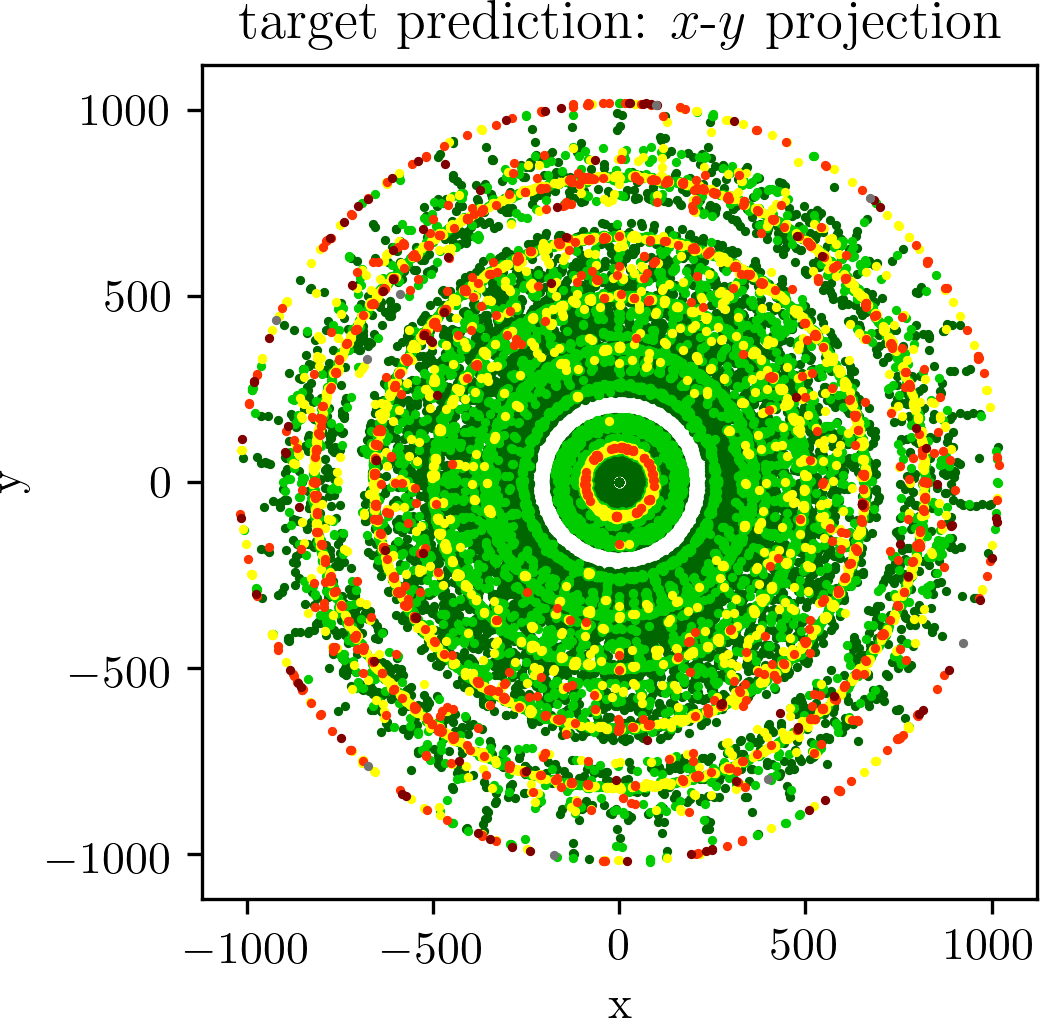}
				\caption{Projection of the results of the target-prediction model onto the $r$-$z$ plane (left) and the $x$-$y$ plane (right). The red dots in the horizontal lines (left) match exactly the overlaps of the surfaces in this area.}
				\label{fig:target_pred_generic_rzmap}
			\end{subfigure}
			\begin{subfigure}[t]{\linewidth}
				\centering
				\includegraphics[height=50mm]{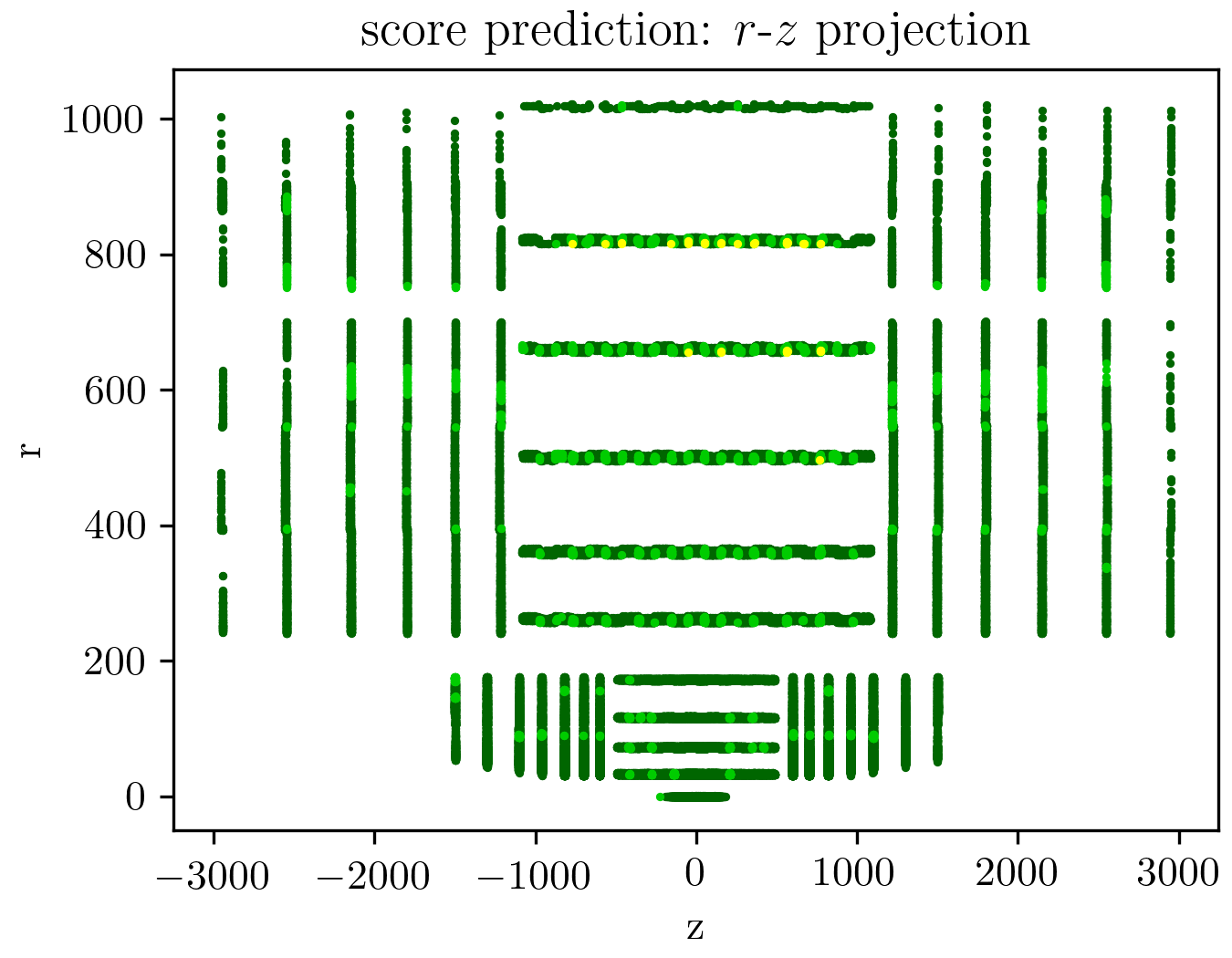}
				\hfill
				\includegraphics[height=50mm]{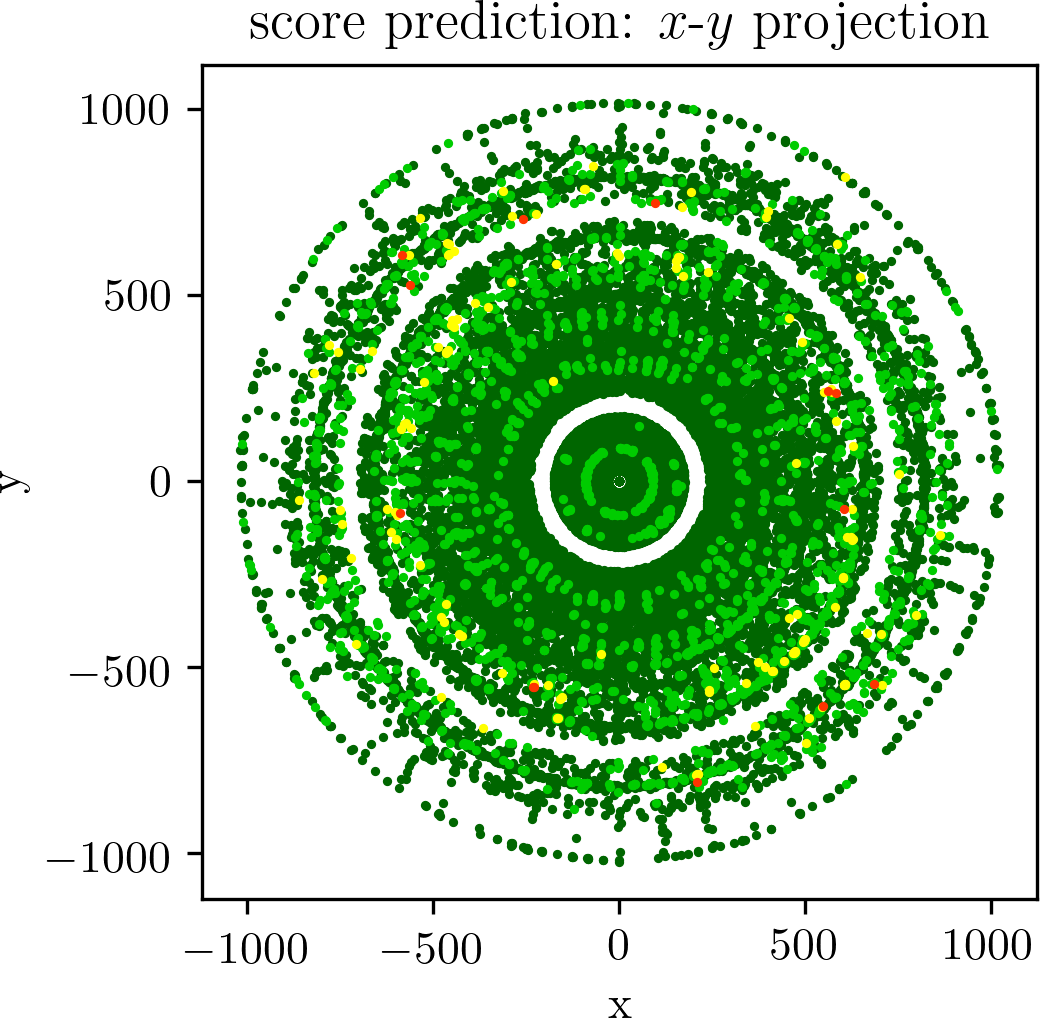}
				\caption{Same as Fig.~\ref{fig:target_pred_generic_rzmap} but for the score-prediction model. As in Fig.~\ref{fig:target_pred_generic_rzmap} (left plot), we can see score fluctuations which correlate with the overlaps. However, the overall performance is much better.}
				\label{fig:pairwise_score_generic_rzmap}
			\end{subfigure}
			\caption{Projection of the score results onto the $r$-$z$ and $x$-$y$ planes. In each case, 100k random points where sampled and replaced by the average of all neighbors (in that plane) within a \SI{5}{\milli\meter} radius. The scores are plotted with better scores in lower plotting layers. Hence better scorses could be hidden by worse scores. The color coding is as in Fig.~\ref{fig:score_distributions}.}
			\label{fig:pairwise_score_data}
		\end{figure}
		
	\subsection{Target prediction}
		For this approach, a pre-trained embedding turned out to be much more effective than the use of the 3D surface coordinates. The use of small beam-pipe splits (e.g., (40,1)) slightly increases the performance. However, even without a beam-pipe split a reasonable performance is obtained. Large beam-pipe splits (e.g., (400,16)) decrease the performance noticeably.
		
		The performance metrics of the target-prediction navigator are evaluated with the help of a KNN search \cite{sklearn}. The ten nearest neighbors are sorted by their distance from the predicted point in the embedding space. If the closest neighbor is also the true target embedding, the prediction is considered to be correct and put it in the score category \emph{correct}. If the true result is one of the other elements, that element is put in the corresponding category (the categories used are \emph{in best 2}, \emph{in best 3}, \emph{in best 5}, \emph{in best 10}). If the correct target does not occur in these 10 points, the prediction is considered to be failed (category \emph{worse}). Results for the generic detector are shown in Figs.~\ref{fig:score_distributions} and \ref{fig:target_pred_generic_rzmap}.
		
		The results show that the target-prediction navigator can reach good performance when using a large enough dataset for training. However, there always remains a small share of cases ($\sim\!5\% $) where the predictions are more than 5 neighbors off. This happens mostly in the barrel sections of the short strip and the long strip and can be correlated to the overlaps of the surfaces in this area (see Fig.~\ref{fig:target_pred_generic_rzmap}, left plot). Further research is needed to show if this can be improved by, e.g., optimizing the embedding.	

		%\afterpage{\clearpage}

	\subsection{Score prediction}
		For the score prediction the most successful approach was to use the center coordinates of the surfaces as the embedding. Additionally, this approach benefits significantly from a fine-grained beam-pipe split.
		The performance metrics of the score-prediction navigator are evaluated by comparing the score of the true target to the score of the predicted target. If the true target has the highest score, the prediction is considered to be correct (category \emph{correct}). If the candidate with the second-highest score is the true target, it belongs to category \emph{in best 2}, etc. (as described in the previous subsection). If the graph only contains a few possible targets, the solution is guaranteed to be found in the, e.g., first 4 targets. This must be taken into account when interpreting the results. 
		
		The problem mentioned earlier, i.e., the possibility that the true target is not present in the graph, cannot occur in the evaluation since the graph is built from the entirety of the available data. In a real simulation environment this problem will occur, in which case a fallback solution must be implemented as mentioned above. Additionally, care must be taken when generating the graph from the data. Graphs generated from data sets like the one described above show non-negligible differences regarding the connections present in the graph. Further studies are needed to understand how to generate the graph in a way that it is less dependent on statistics.
		
		The performance of the score-prediction approach, see Figs.~\ref{fig:score_distributions} and \ref{fig:pairwise_score_generic_rzmap}, is notably better than that of the target prediction, especially in the barrel regions. Most importantly, in nearly 100\% of the cases the prediction lies within the first 5 candidates. This, together with the fact that no nearest-neighbor search must be performed to resolve the targets, makes it the preferred approach for ML-based navigation.

\section{Implementation in ACTS}
	% The plan for a proof-of-concept implementation in ACTS consists of two parts: first, the score-prediction navigator tries to predict the next surface, based on a graph with possible target surfaces. The navigator then tries to intersect the track with the candidate surfaces in order of their score. If that fails (i.e., if the track cannot intersect with any candidate), a fallback navigator on the basis of the target-prediction model is used.
	
	%To load the neural networks into ACTS, the ONNX API and runtime \cite{onnx} will be used. A (not yet feature-complete) version of the ACTS implementation can be found at \cite{my_github_repo}. 
	
	We provide a proof-of-concept implementation that comprises both discussed models, the score-prediction navigator and the target-prediction navigator, at \cite{my_acts_fork}. To load the neural networks into ACTS, the ONNX API and runtime \cite{onnx} is used.

\section{Conclusion and Outlook}
	The results show that it is possible to use neural networks for navigation. Furthermore, these approaches can be applied to almost any detector geometry in combination with a trial-and-error navigator. It is expected that with more extensive hyperparameter optimization the results can improve further.
	
	No detailed analysis regarding the runtime performance, the memory consumption and the computational overhead of the different approaches has been done yet. However, first tests show that the neural network inference with the current models is about three orders of magnitude slower than the straight-line intersection (which is used to verify a surface hit during integration). In order to make this approach computationally competitive, the runtime of the neural network inference must be improved substantially, e.g., by simplifying the network architecture or by large-scale parallelization. This is the subject of future work. 
	
	%, hence no statement can be made regarding whether these approaches will eventually match the requirements of all possible application scenarios. However, 
	
	Regarding further research, not all potential model architectures have been tried for this task yet. For example, the potential of recurrent neural networks is yet completely unexplored. This will also be the subject of further research.

\addcontentsline{toc}{section}{References}
\bibliography{bibliography}

\end{document}